**Qualitative Assessment of Gene Expression in Affymetrix Genechip Arrays**

Radhakrishnan Nagarajan[*]

Institute on Aging, University of Arkansas for Medical Sciences

Meenakshi Upreti
Department of Biochemistry and Molecular Biology, University of Arkansas for Medical

Sciences

To whom correspondence should be addressed:

Radhakrishnan Nagarajan
Institute on Aging
University of Arkansas for Medical Sciences
629 Jack Stephens Drive, Room: 3105
Little Rock, AR 72205, USA
Email: nagarajanradhakrish@uams.edu

**Abstract**

Affymetrix Genechip microarrays are used widely to determine the simultaneous expression of genes in a given biological paradigm. Probes on the Genechip array are atomic entities which by definition are randomly distributed across the array and in turn govern the gene expression. In the present study, we make several interesting observations. We show that there is considerable correlation between the probe intensities across the array which defy the independence assumption. While the mechanism behind such correlations is unclear, we show that scaling behavior and the profiles of perfect match (PM) as well as mismatch (MM) probes are similar and immune to background subtraction. We believe that the observed correlations are possibly an outcome of inherent non-stationarities or patchiness in the array devoid of biological significance. This is demonstrated by inspecting their scaling behavior and profiles of the PM and MM probe intensities obtained from publicly available Genechip arrays from three eukaryotic genomes, namely: Drosophila Melanogaster (fruit fly), Homo Sapiens (humans) and Mus musculus (house mouse) across distinct biological paradigms and across laboratories, with and without background subtraction. The fluctuation functions were estimated using detrended fluctuation analysis (DFA) with fourth order polynomial detrending. The results presented in this study provide new insights into correlation signatures of PM and MM probe intensities and suggests the choice of DFA as a tool for qualitative assessment of Affymetrix Genechip microarrays prior to their analysis. A more detailed investigation is necessary in order to understand the source of these correlations.

## 1. Introduction

Affymetrix Genechip microarrays [1-3] have been used widely to determine the simultaneous expression of a large number of genes in distinct biological paradigms. Several algorithms have been proposed in the past to determine differential gene expression across distinct biological states [1-2], model gene interactions [4-6] and classify pathological conditions [7]. There have been reports in the past that investigated the existence of spurious spatial bias in log-transformed gene expression values [8] obtained from microarrays. These studies also demonstrated characteristic pattern in gene expression as a function of chromosomal distance. However, the gene expression of a transcript in Affymetrix Genechip arrays is estimated from their probe intensities, which by very design are spotted randomly on a Genechip with no apparent pattern. Therefore, the focus of the present study is on the probe intensities which are the atomic elements that govern gene expression. Gene expression of a transcript is estimated as complex combination of these atomic probe intensities [3]. Restricting the analysis at the atomic level also prevents any possible correlations that might be introduced by the gene expression estimation procedure. As noted earlier, the Genechip in essence is a *random matrix* whose elements are uncorrelated. In the present study we show evidence of considerable correlations in probe intensities across publicly available Genechip arrays from three eukaryotic genomes namely: Drosophila Melanogaster (fruit fly), Homo Sapiens (humans) and Mus musculus (house mouse), across laboratories [9] and across biological paradigms (Table I). This is accomplished by inspecting the probe intensity profiles along with their scaling behavior using (DFA) [9] with fourth order polynomial detrending [10, 11]. The choice of analyzing Genechips across organisms is to reject the claim that the observed correlation is an outcome of layout of a specific Genechip. Analyzing Genechip arrays across paradigms and laboratories [9] minimizes the possibility that the observed correlation is an outcome of experimental protocols of a laboratory or a specific paradigm. Background subtraction [12, 13] is an important pre-processing step in gene expression analysis and minimizes the effect of non-specific hybridization which in turn can contribute to spatial bias.

In the present study we compare the PM and MM probe intensities before and after background subtraction in order to reject the claim that the observed correlations are an outcome of background subtraction. While PM is a measure of specific-binding, MM is a measure of non-specific binding and its role in estimating gene expression has remained elusive [12-15]. Subsequently PM only models have been proposed for gene expression estimates [13]. In spite of the discrepancy in their binding efficiencies we show that both the PM and the MM intensities exhibit similar correlation signatures across all the Genechip arrays considered. While the cause of the observed correlations is unclear, we believe they're non-biological and an outcome of inherent non-stationarities or patchiness in the probe intensity data. The results presented raise fundamental questions on the interpretation of gene expression data. For readability and completeness we first review some of the essential nomenclature of Affymetrix Genechip arrays [1] prior to discussion of the probe intensity data.

*Affymetrix Genechip Arrays*

The atomic entity of the Affymetrix Genechip array [1] is a *probe* (e.g. 5'-GTGATCGTTTACTTC GGTGCCACCT-3') distributed randomly across the array. A probe is usually (~ 25 nucleotides long). A set of (~16 to 20) probes also called a *probeset*, represents a particular transcript or a gene on the array. The term transcript is generic and can represent a gene or an expressed sequence tag (EST). Probes can be further classified into *perfect match* (PM) or *mismatch* (MM) probes. The PM probes (e.g. PM: 5'-GTGATCGTTTACTTCGGTGCCACCT-3') correspond to a short region of the transcript and are designed to be complementary to the *target* (5'-CACTAGCAAATGAAGCCACGGTGGA-3'). The nucleotide content of an MM probe is the same as that of the corresponding PM probe except for the middle most nucleotide, which is flipped deliberately (e.g. MM probe corresponding to the PM probe shown above is: 5'-GTGATCGTTTACTCCGGTGCCACCT-3'). Thus the MM probe is used an internal control to assess non-specific (non-biological) hybridization and are located physically adjacent to the corresponding PM probe on the array, Fig. 1. For the same reason, one often refers to the (PM, MM) probes as *probe pairs*. An important feature of the Genechip array is that the (PM, MM) probe pairs corresponding to a transcript are distributed *randomly* across the array with no apparent pattern. Thus ideally, the entire array can be regarded as a random matrix with PM and MM probes distributed in pairs. The location of the 14 (PM, MM) probe pairs for the transcript 145795_at (Drosophila Genechip Array, Table I) is shown in Fig. 2a. The PM and MM intensities across distinct paradigms (D1, D2, Table I) and across labs (Lab 1, Lab 2, Table I) for the 14 (PM, MM) probe pairs is *non-uniformly* distributed. Such non-uniform distribution is to be expected as the nucleotide content of the PM and MM probes corresponding to a given transcript need not necessarily be the same which in turn affects their binding efficiency to a given target. Recent reports have implicated such probe-to-probe variations to variations in the hybridization free energies [15]. From Fig. 2, the PM and MM probe intensities also show considerable variation for the given probe set (145795_at) across distinct paradigms (D1, D3, Table I) and across labs. For the above reasons, a generic form of the distribution is usually absent. This in turn implies extension of the results at the level of probe intensities to that of probeset is neither trivial nor straightforward.

## 2. Methods

In a typical Genechip microarray experiment, tissues from a given specimen (e.g. tumor specimen) are hybridized onto the array. Hybridization is a complex procedure with several intermediate steps [1-3]. Subsequently, these arrays are washed, stained and laser scanned at a particular wavelength to yield the image files (.DAT files), which in turn yield the probe intensities, stored in (.CEL files). As noted earlier the probes corresponding to a probeset or transcript, are placed randomly across the array. The location of the probes is specified by the (.CDF file). In the present study, we use the .CEL file in conjunction with the .CDF file to extract the location and intensity of the PM and MM probes. Since the objective of the study to investigate presence of possible correlations, we map the two-dimensional matrix of PM and MM intensities into a one-dimensional vector of PM and MM intensities. A schematic diagram explaining the mapping procedure is shown in Fig. 1. Since the position of the probe intensities on array correspond to time-scale on the one-dimensional data, we shall use the terms position and time scales interchangeably in the subsequent discussions.

Affymetrix Genechip array by their very design have certain markers on the chip these correspond to zero probe intensities. These markers are a part of the chip design and are chip specific. The percentage of such zero probe intensities was quite low across the three Genechip arrays were (i) Rattus Norvegicus (~ 0.8%), Mus musculus (~ 2.7%), Homo Sapiens (~0.85%) and Drosophila Melanogaster (~ 0.9%). These low numbers are unlikely to have any significant impact on the subsequent discussion. Nevertheless, we imposed uncorrelated structure for these probe intensities from random samples from lognormally distributed uncorrelated noise whose mean and variance were determined by the non-zero probe intensities on the array. Static, memoryless transforms such as log-transform has been used widely in microarray gene expression community in order to accommodate near-normality assumptions [12, 13], hence the choice of lognormal distribution. Background subtraction is also encouraged in microarray literature as an important pre-processing step in order to minimize the effect of systematic spatial drift across the array. However, the choice of algorithm to correct the background is an area of active research. Two popular algorithms used widely are the Bioconductor [16] implementation of Affymetrix proprietary algorithm (MAS 5.0) [1, 16] and the more recent robust multichip average (RMA) [13, 16]. These algorithms are publicly available [16] and their details can be found elsewhere [1, 13]. We shall refer to these algorithms as MAS and RMA in the subsequent sections. The qualitative behavior of the raw PM and MM intensities is compared to those obtained by subtracting their respective backgrounds using MAS and RMA. Such an approach eliminates the possibility that the observed correlations are an outcome of varying background across the Genechip.

## 3. Results

Prior to investigating the correlation aspects we investigated the distribution of the PM and MM intensities. Recent studies [17] have provided overwhelming evidence of power-law scaling of the form $P(k) \sim k^{-g}$ of the distribution of gene expression values across several Genechip arrays and cDNA arrays [18, 19], where $k$ represents the magnitude of the gene expression and $P(k)$ represents the frequency of its occurrence. Such power-law behavior had been attributed to universality in transcriptional organization across genomes in [17]. As noted earlier, extension of the results obtained on the probeset intensities to those at probe intensities is not immediate. This can be attributed ti the fact that the probeset intensities are derived as a complex combination of probe intensities [1, 12, 13]. Surprisingly, we found such power-law decay of the distribution to hold even at the level of probe intensities. More importantly, the power-law decay persisted across the PM as well as MM values. It is important to recall that PM represents specific binding whereas MM measures non-specific binding. The power-law decay also persisted across the three eukaryotic genomes, across paradigms, across labs and across the raw and background subtracted PM and MM intensities. The log-log plot of the frequency of occurrence against that of the magnitude of expression for the one-dimensional PM and MM intensities before and after background subtraction across the arrays (Table I) is shown in Fig. 3 with the PM intensities being considerably larger than that of MM, reflected by the upward shift in the slope of the curve corresponding to PM intensities, Fig. 3. The exponents of the MM intensities across the arrays (D1, D2, D3, $g^D$ ), (H1, H2, H3, $g^H$ ) and (M1, M2, M3, $g^M$ ), Table I, were in the range $g^D \in (1.6 \text{ to } 2.78)$ , $g^H \in (2.1 \text{ to } 2.4)$ and $g^M \in (2.3 \text{ to } 3.4)$ respectively. A similar analysis revealed the PM intensities for the above cases were $g^D \in$ (1.3 to 3.1), $g^H \in$ (2.2 to 2.8) and $g^M \in$ (2.3 to 2.5). In the following discussion we show that in addition to the power-law distribution, the PM and MM intensities also exhibit similar correlation signatures across the above Genechip arrays.

The correlation of the one-dimensional PM and MM intensities was investigated using DFA with fourth order polynomial detrending. The choice of higher order polynomial detrending is attributed to a recent study [11], which showed $p^{th}$ polynomial detrending of the profile is useful in minimizing local polynomial trends up to order ($p$-1) in the given data. Since the objective is to understand the variation in the intensities across the entire array, we also investigate the profiles [10] in conjunction to their scaling behavior. This can be attributed to the fact that two data sets with distinct profiles can exhibit same scaling behavior. A classic example is that of a monofractal Gaussian noise with exponent ($\alpha = 0.8$) generated using Makse's algorithm [20] and its phase-randomized surrogate (FT) counterpart [21], Fig. 4. FT surrogates are constrained randomized realizations, where the constraint is on retaining the auto-correlation function of the empirical sample in the surrogate realization. By definition FT surrogates retain the correlation function of monofractal Gaussian noise [21], hence their scaling exponent [22], Fig. 4b. However, their profiles can be quite different, Fig. 4a. Also, shown in Fig. 3 is the profile and scaling of the random shuffled counterpart [21] ($\alpha = 0.5$) of the monofractal Gaussian noise which by definition retains the probability distribution and not the correlation. In following discussion we identify the random shuffled counterpart to scaling of the PM and MM intensities obtained by random-parsing of the Genechip matrix.

The plot of the fluctuation $\log_2 F(s)$ versus position $\log_2(s)$ for the PM and MM intensity values of the Drosophila Melanogaster Genechip (D1, Table I) before and after background subtraction obtained by row-wise parsing is shown in Fig. 5 and reveals characteristic crossovers with three different scaling regimes for the MM as well as PM intensities, Figs. 5a and 5b. Crossovers render the scaling of $\log_2 F(s)$ versus $\log_2(s)$ nonlinear. Therefore, linear regression of $\log_2 F(s)$ vs $\log_2(s)$ across the entire length of (s) in the presence of crossovers can result in spurious results. In order to circumvent these issues we estimated the local scaling exponent $\alpha(s)$, by linear regression of overlapping moving windows. This was accomplished by choosing a window containing five points, estimate the exponent by local linear regression of the points in that window $\alpha(s)$, shift the window by two points and repeat the procedure. As a result, we obtain the local scaling exponents $\alpha(s)$ as a function of the position $\log_2(s)$, Figs. 5c and 5d. Closer inspection of the local slopes reveals pronounced crossovers with three distinct scaling regimes for PM as well as MM intensities, obtained by row-wise parsing prior to and after background subtraction (MAS, RMA) [12, 13, 16]. Since background subtraction has negligible effect on the scaling one fails to see separate curves. The three distinct scaling regimes correspond to uncorrelated, correlated and anti-correlated behavior ($\alpha 1 \sim 0.5$, $\alpha 2 > 0.5$, $\alpha 3 < 0.5$) in time-scales (positions) $s \in (2^6, 2^7)$, $s \in (2^7, 2^{9.5})$, $s \in (2^{9.5}, 2^{10.5})$. The noise in the fluctuation function increases for $(s > 2^{10.5})$, hence any conclusions at these time-scales are unreliable, Figs. 5a and 5b. For the same reason, we shall restrict the discussion in the subsequent sections only for time-scales $(s < 2^{10.5})$. As expected, the scaling of the PM and MM intensities obtained by random-parsing, Fig. 1, fails to exhibit any characteristic crossovers and local scaling exponents is close to ($\alpha 2 \sim 0.5$) in the time-scales $s \in (2^6, 2^{10.5})$. Thus from the above discussion it is clear that the PM and MM intensities exhibit similar correlation signatures which persists across background subtraction. While it is tempting to attribute the above correlation in the PM and MM intensities to interesting dynamics, we believe they're solely an outcome of non-stationarities possibly due to experimental artifacts inherent to the microarrays.

A similar analysis of the raw and background subtracted (MAS, RMA) PM and MM intensities from three eukaryotic Genechip arrays from Drosophila Melanogaster (D1, D2, D3), Homo Sapiens (H1, H2, H3), Mus musculus (M1, M2, M3) across laboratories and across experiments, Table I is shown in Fig. 6 The corresponding *profiles* [10] were generated as partial sums (integrated series) of the mean subtracted (PM, MM) intensities as a function of their position. Since the PM and MM intensities differ significantly in their magnitude across the arrays, the profiles were normalized to zero-mean unit variance to facilitate qualitative comparison. In the subsequent discussion, profile implicitly refers to normalized profile. The profile for the various eukaryotic Genechip arrays is shown in Fig. 7. Genechips suffixed with (1, 2, e.g. D1, D2) were chosen from the same lab where those suffixed with (3, e.g. D3) were chosen from a different lab, Table I. As observed earlier, Fig. 5, the Genechip arrays of Drosophila Melanogaster (D1, D2, D3) exhibits three distinct scaling regimes, i.e. uncorrelated to correlated to anti-correlated in the time-scales $s \in (2^6, 2^7)$, $s \in (2^7, 2^{9.5})$, $s \in (2^{9.5}, 2^{10.5})$, Figs. 6a, 6d and 6g. The scaling of the MM intensities follows that of the PM intensities reflected by their significant overlap. The corresponding profile of the raw and background subtracted PM and MM intensities are shown in Figs. 7a, 7d and 7g. Qualitative inspection of the profiles for arrays printed within a lab (D1, D2, Lab 1, Table I) reveal higher degree of similarity than those printed across labs (D3, Lab 2). A similar analysis of the raw and background subtracted PM and MM intensities of Homo Sapiens (H1, H2, H3, Table I) is shown in Figs. 6b, 6e, 6h, and Figs. 7b, 7e, 7h. The fluctuation function exhibits three distinct scaling regions, i.e. uncorrelated to correlated to anti-correlated in the time-scales $s \in (2^6, 2^9)$, $s \in (2^9, 2^{9.5})$, $s \in (2^{9.5}, 2^{10.5})$. However, the correlated regime corresponding to time-scales $s \in (2^9, 2^{9.5})$ is less prominent than in the case of Drosophila Melanogaster. The profiles for arrays printed within a lab (H1, H2, Lab 3, Table I) reveal higher degree of similarity than those printed across labs (H3, Lab 4). Investigating the fluctuation function of raw and background subtracted PM and MM intensities from Mus Musclus Genechip arrays, Figs. 6c, 6f and 6i, revealed scaling behavior similar to those of Homo Sapiens. While the fluctuation function were similar across (M1, M2, M3), Figs. 7c, 7f and 7i, the profiles exhibited considerable discrepancy for arrays printed across labs (M3) than those printed within the same lab (M1, M2). It is important to note in the above discussion, the fluctuation functions, Fig. 6 and the profiles Fig. 7, of the raw and background subtracted (MAS 5.0, RMA) PM and MM intensities failed to show any significant discrepancies. This was true across paradigms and across labs. Alternately, background subtraction cannot explain the observed correlation.

## 4. Discussion

Affymetrix Genechip microarrays have been used widely to determine the simultaneous expression of a large number of genes in biological paradigms. Developing novel techniques for interpreting gene expression is an area of active research. In the present study we employ tools of statistical physics to gain insight into gene expression at the level of probe intensities. While such an approach is unconventional, it nevertheless provides new insight into the nature of correlations in Genechip probe intensities which to our knowledge has never been investigated. The probes are spotted on a Genechip array in a random fashion. In essence, the Genechip array is a random matrix with uncorrelated elements. This possibly has encouraged the choice of statistical tests that infer gene expression under implicit independence assumption of the probe intensities. In the present study, we first mapped the two-dimensional matrix of PM and MM intensities into one-dimensional vectors by row-wise and random-parsing. We showed that a systematic row-wise parsing reveals correlation at distinct time scales in sharp contrast to those of random parsing. Such correlations were demonstrated across PM and MM intensities from three eukaryotic Genechip arrays across labs, across paradigms, with and without background subtraction. While PM is a measure of specific binding, MM is a measure of non-specific binding used as an internal control. Understanding the behavior of MM probes is still a mystery in the microarray research, subsequently PM only models have been proposed to infer differential gene expression.

Power-law distributed gene expression signatures in Affymetrix Genechip arrays were attributed to universality in transcriptional organization across genomes. Such power-law distributions have also been observed in the case of cDNA arrays and subsequently used to understand the underlying network structure [22]. In the present study, we found that such power-law distribution to persist even at the atomic level (PM, MM). In Affymetrix Genechip arrays, gene expression is estimated as a complex combination of PM and/or MM intensities. Therefore, extension of the results at the level of gene expression to those of probe intensities is neither-trivial or straightforward. Interestingly in the present study, we also found PM and MM intensities exhibited similar correlation signatures revealed by their scaling and profiles. While the scaling exhibited considerable deviation from ($\alpha = 0.5$), the profiles exhibited large excursions unlike their random shuffled counterpart. Background subtraction is an important pre-processing step and encouraged to minimize drift across the array due to non-specific hybridization. However, the results presented were immune to background subtraction accomplished using two popular algorithms (RMA, MAS). Qualitative inspection of the profile revealed that the arrays from the same lab are likely to be more similar than arrays across different labs. The results presented raise an important question whether the observed correlation can be due to inherent non-stationarities devoid of biological significance. This is more so, since the correlation signatures persists across the PM and the MM probe intensities. Clustering techniques along with correlation metrics are used widely in microarray gene expression analysis to infer functional dependencies.

From the perspective of the present study, cautious interpretation results based on correlation metrics is necessary in order to avoid false-positives. The present study also encourages the investigating the fluctuation function and the profiles of the probe intensity data as qualitative inspection of the Genechip arrays prior to their analysis. In a parallel study on cDNA microarrays, power-law distribution of the gene expression was used to identify the underlying network structure [22].

## Acknowledgements

We would like to thank NCBI GEO and the contributors of the data sets listed in Table I for generously making available their experimental data sets.

23. Sources for the data sets listed under Table I

**Mus Musculus:**      Data: M1, M2, M3 Source: http://www.ncbi.nlm.nih.gov/geo/
**Homo Sapiens:**       Data: H1, H2, H3 Source: http://www.ncbi.nlm.nih.gov/geo/
**Drosophila  Melanogaster:**
                   Data D1, D2 Source: http://www.fruitfly.org/expression/immunity/
                   Data: D3 Source: http://www.ncbi.nlm.nih.gov/geo/

Given microarray gene expression matrix with probe intensities $PR_{i,j}, i = 1...r, j = 1...c$ with $r = 4$, $c = 4$. PM and MM probes occur in pairs (probe pairs) adjacent to each other (e.g. $PR_{11}$, $PR_{12}$) and are shown in white and gray respectively.

| $PR_{11}$ | $PR_{12}$ | $PR_{13}$ | $PR_{14}$ |
|-----------|-----------|-----------|-----------|
| $PR_{21}$ | $PR_{22}$ | $PR_{23}$ | $PR_{24}$ |
| $PR_{31}$ | $PR_{32}$ | $PR_{33}$ | $PR_{34}$ |
| $PR_{41}$ | $PR_{42}$ | $PR_{43}$ | $PR_{44}$ |

Mapping of the above gene expression matrix into one-dimensional vectors PM and MM probe intensities by:

(a) Row-wise parsing

PM: $PM_{2(i-1)+j} = PR_{i,2j-1}, i = 1...r; j = 1...c/2$

| $PM_1$ | $PM_2$ | $PM_3$ | $PM_4$ | $PM_5$ | $PM_6$ | $PM_7$ | $PM_8$ |
|--------|--------|--------|--------|--------|--------|--------|--------|
| $PR_{11}$ | $PR_{13}$ | $PR_{21}$ | $PR_{23}$ | $PR_{31}$ | $PR_{33}$ | $PR_{41}$ | $PR_{43}$ |

MM: $MM_{2(i-1)+j} = PR_{i,2j}, i = 1...r; j = 1...c/2$

| $MM_1$ | $MM_2$ | $MM_3$ | $MM_4$ | $MM_5$ | $MM_6$ | $MM_7$ | $MM_8$ |
|--------|--------|--------|--------|--------|--------|--------|--------|
| $PR_{12}$ | $PR_{14}$ | $PR_{22}$ | $PR_{24}$ | $PR_{32}$ | $PR_{34}$ | $PR_{42}$ | $PR_{44}$ |

(a) Random parsing

PM*: random shuffle of the one-dimensional vector PM

| $PM*_1$ | $PM*_2$ | $PM*_3$ | $PM*_4$ | $PM*_5$ | $PM*_6$ | $PM*_7$ | $PM*_8$ |
|--------|--------|--------|--------|--------|--------|--------|--------|
| $PR_{21}$ | $PR_{31}$ | $PR_{11}$ | $PR_{41}$ | $PR_{33}$ | $PR_{13}$ | $PR_{43}$ | $PR_{23}$ |

MM*: random shuffle of the one-dimensional vector MM

| $MM*_1$ | $MM*_2$ | $MM*_3$ | $MM*_4$ | $MM*_5$ | $MM*_6$ | $MM*_7$ | $MM*_8$ |
|--------|--------|--------|--------|--------|--------|--------|--------|
| $PR_{22}$ | $PR_{32}$ | $PR_{12}$ | $PR_{42}$ | $PR_{34}$ | $PR_{14}$ | $PR_{44}$ | $PR_{24}$ |

**Figure 1** Mapping of the PM and MM probes on the Affymetrix Genechip array matrix into one-dimensional vectors by row-wise parsing (PM and MM) and random parsing (PM* and MM*).

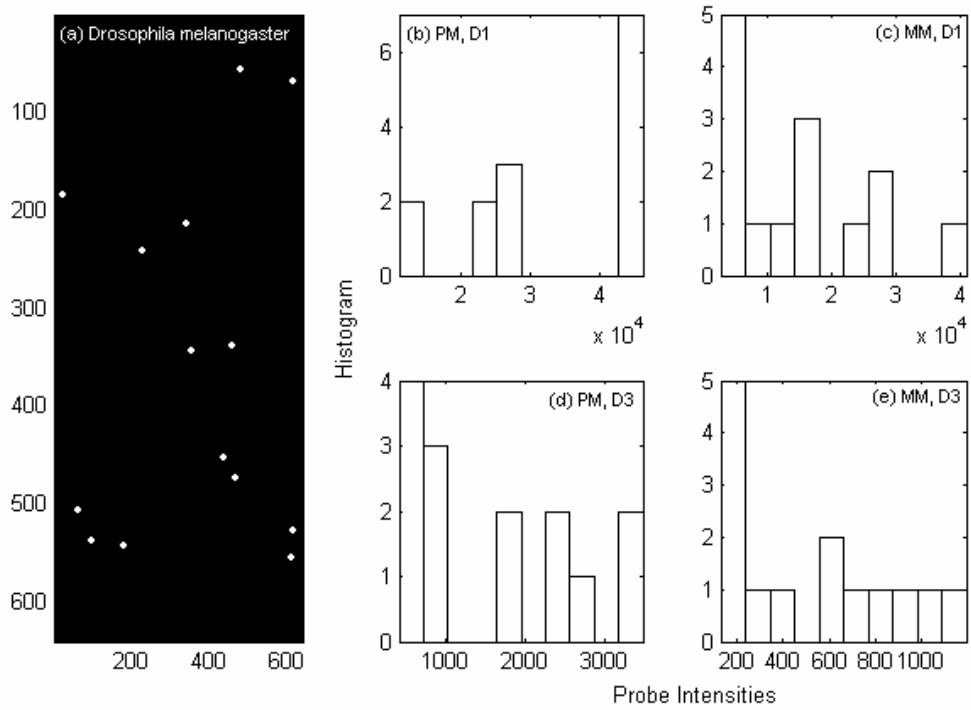

**Figure 2** Spatial location of the $PM_n$ and $MM_n$ ($n = 1\ldots14$) probe intensities corresponding to the probeset (l45795_at) in Drosophila Melanogaster Genechip array (a). Non-uniform distribution of the PM (b, d) and MM (c, e) probe intensities corresponding to the probeset (145795_at) across distinct biological states (D1, Lab 1, Table I), (b), (c) and across laboratories (D3, Lab 2, Table I), (d), (e).

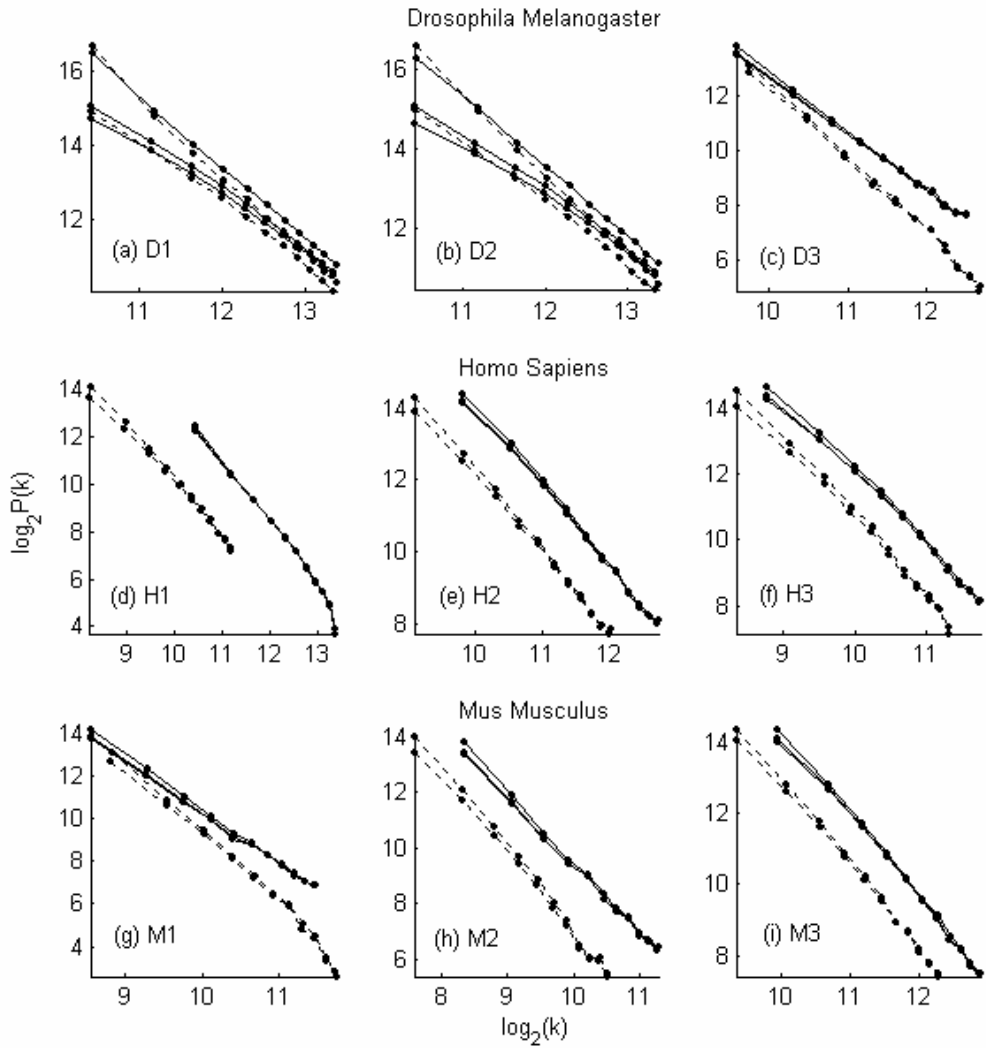

**Figure 3** Variation in the magnitude of the one-dimensional vector of PM (solid lines) and MM (dotted lines) probe intensities $\log_2 k$ against their probability of occurrence $\log_2 P(k)$ across distinct biological states and laboratories (Table I) for Drosophila (D1, D2, D3, Table I), Homo Sapiens (H1, H2, H3, Table I) and Mus Musculus (M1, M2, M3, Table I) Genechip arrays with and without background subtraction.

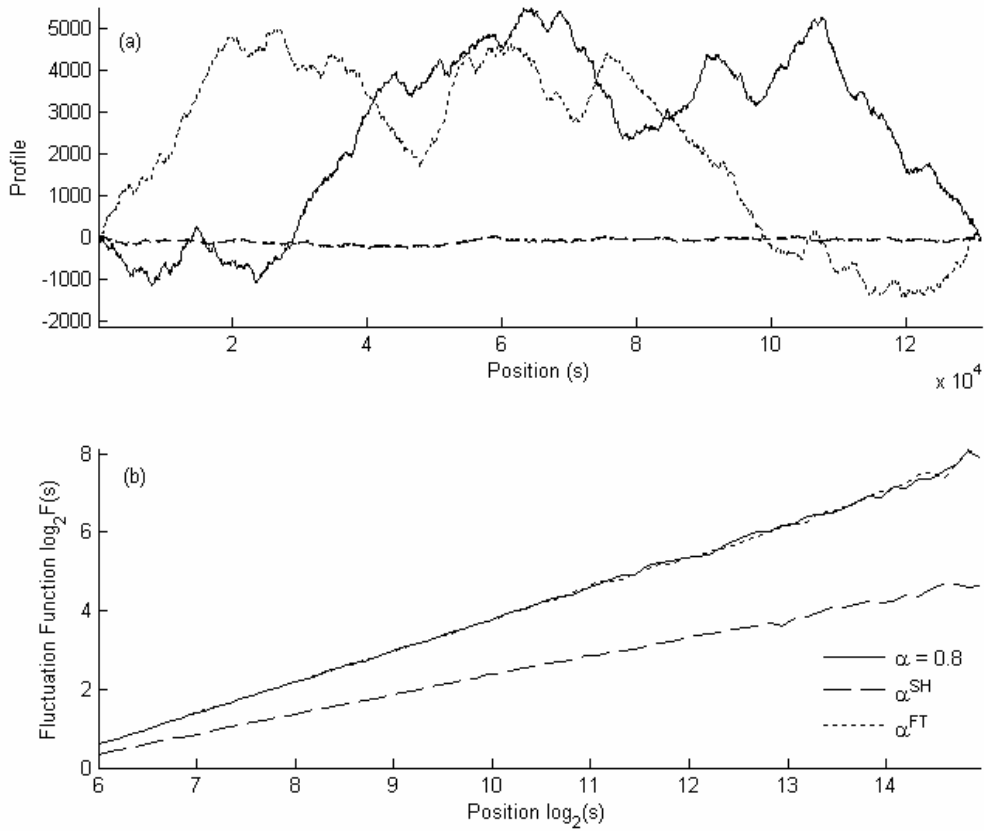

**Figure 4** Profiles of the monofractal Gaussian noise with scaling exponent ($\alpha = 0.8$, solid line) and its corresponding phase-randomized ($\alpha^{FT} = 0.8$, dotted line) and random shuffled ($\alpha^{SH} = 0.5$, dashed lines) surrogates is shown in (a). Plot of the fluctuation function $\log_2 F(s)$ versus timescale $\log_2(s)$ for the three cases in (a) obtained using DFA with fourth order polynomial detrending (b).

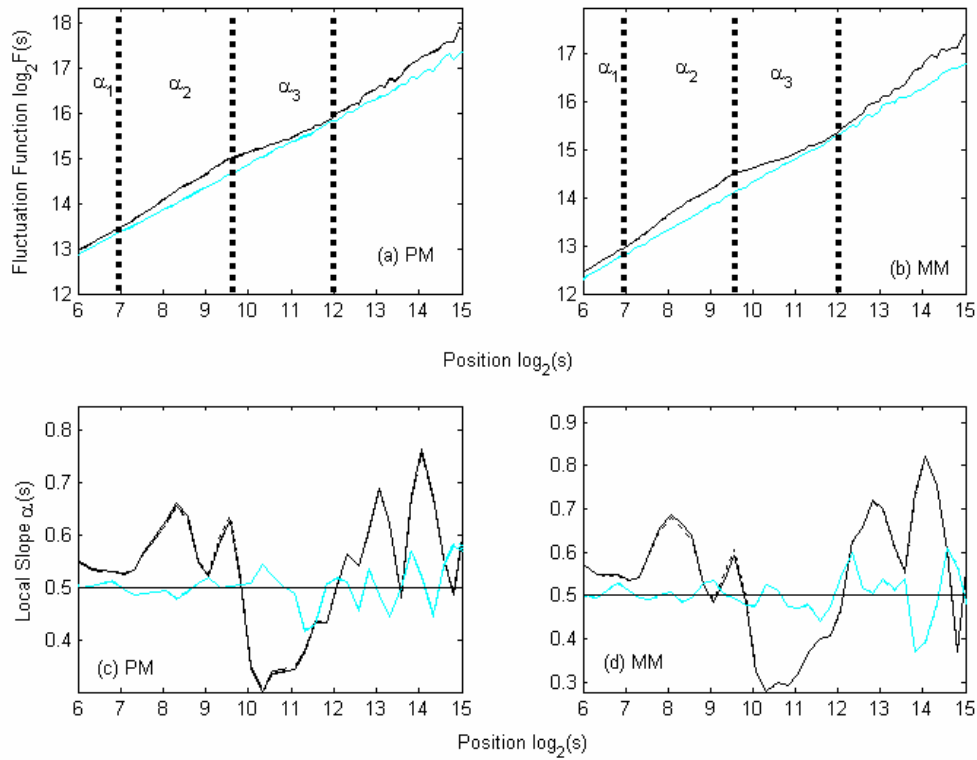

**Figure 5** Plot of the fluctuation function $\log_2 F(s)$ versus timescale (position) $\log_2(s)$ of the PM (a) and MM (b) intensities obtained by row-wise parsing (dark color) and random parsing (light color) of Drosophila Melanogaster Genechip array (D1, Table I) using DFA with fourth order polynomial detrending. The probe intensities obtained with and without background subtraction (MAS, RMA) are shown by (solid lines, dotted lines and dashed lines) respectively in the subplots. The vertical solid lines in (a) and (b) separate three different scaling regimes ($\alpha 1$, $\alpha 2$ and $\alpha 3$) corresponding to uncorrelated, correlated and anti-correlated behavior. The local slopes of the fluctuation functions corresponding to (a) and (b) are shown right below them in (c) and (d) respectively.

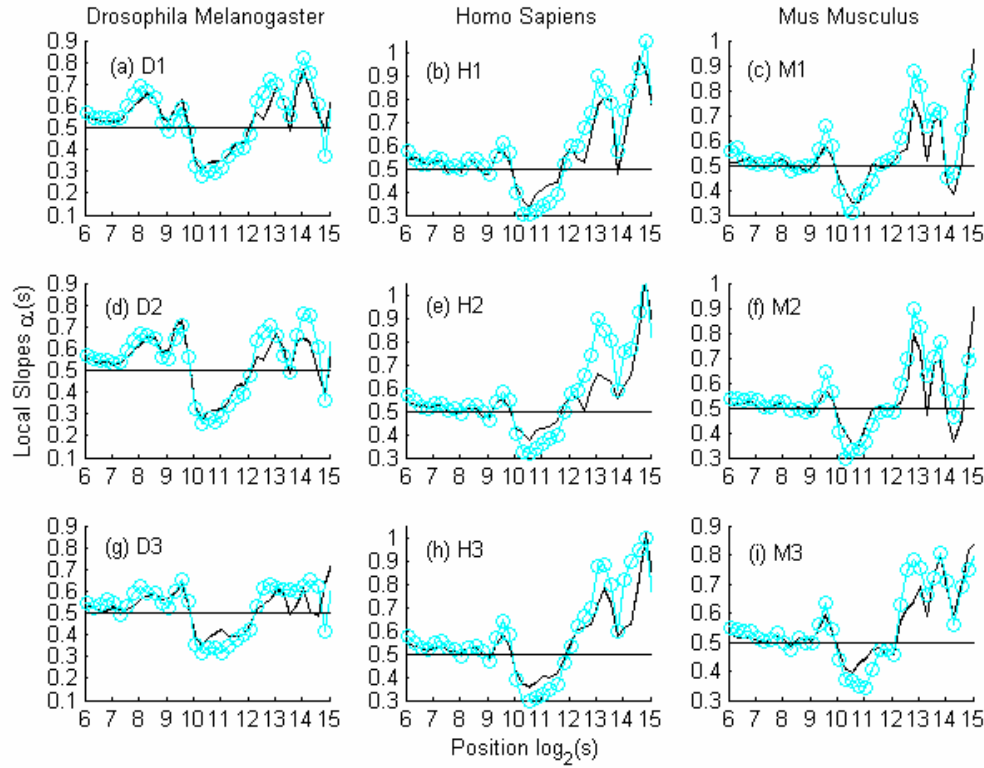

**Figure 6** Plot of the local scaling exponents α(s) versus position log₂(s) obtained by DFA with fourth order polynomial detrending of the raw PM (dark colored lines) and MM intensities (light colored lines with circles) obtained by row-wise parsing of Drosophila Melanogaster (D1 (a), D2 (d), D3 (g), Table I), Homo Sapiens (H1 (b), H2 (e), H3 (h), Table I) and Mus Musculus (M1 (c), M2 (f), M3 (i), Table I) Genechip arrays across distinct biological states and laboratories. The local scaling exponents of the PM and MM intensities after background subtraction using MAS (dotted lines), RMA (dashed lines) are also enclosed in the respective subplots. The horizontal solid lines corresponds to uncorrelated noise obtained by random parsing (α = 0.5) and is shown as a reference.

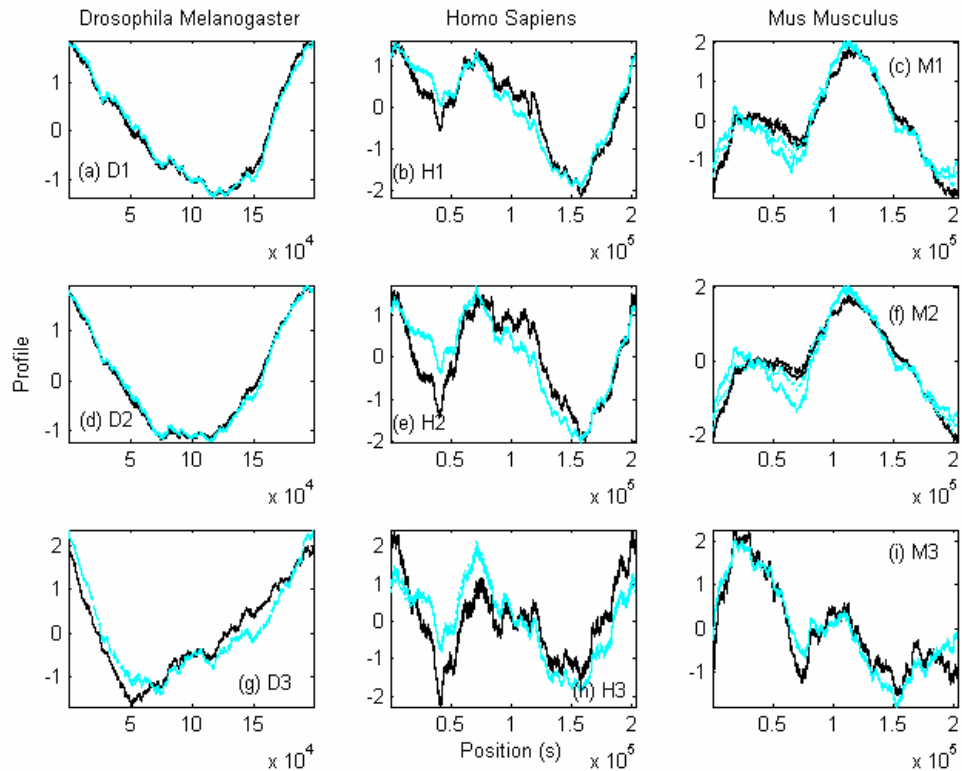

**Figure 7** Plot of the normalized profiles versus position (s) of the raw PM (dark colored lines) and MM intensities (light colored lines) obtained by row-wise parsing of Drosophila Melanogaster (D1 (a), D2 (d), D3 (g), Table I), Homo Sapiens (H1 (b), H2 (e), H3 (h), Table I) and Mus Musculus (M1 (c), M2 (f), M3 (i), Table I) Genechip arrays across distinct biological states and laboratories. The profiles of the PM and MM intensities after background subtraction using MAS (dotted lines), RMA (dashed lines) are also enclosed in the respective subplots.

**Table I**

Affymetrix Genechip microarrays from three eukaryotic genomes, across distinct experimental paradigms and laboratories (see Reference 23)

| Organism | GEO Platform ID | Affymetrix Genechip Name | Labs | ID | Accession Number |
|---|---|---|---|---|---|
| Drosophila | GPL72 | DrosGenome1 | Lab 1 | D1 | s8_20010417 |
| Drosophila | GPL72 | DrosGenome1 | Lab 1 | D2 | s16_20010417 |
| Drosophila | GPL72 | DrosGenome1 | Lab 2 | D3 | GSM29173 |
| Homo Sapiens | GPL91 | HGU95Av2 | Lab 3 | H1 | GSM23162 |
| Homo Sapiens | GPL91 | HGU95Av2 | Lab 3 | H2 | GSM23185 |
| Homo Sapiens | GPL91 | HGU95Av2 | Lab 4 | H3 | GSM4843 |
| Mus Musclus | GPL81 | MGU74Av2 | Lab 5 | M1 | GSM6072 |
| Mus Musclus | GPL81 | MGU74Av2 | Lab 5 | M2 | GSM6073 |
| Mus Musclus | GPL81 | MGU74Av2 | Lab 6 | M3 | GSM2340 |